\documentclass[prd,showkeys,floatfix,nofootinbib,
               preprint,12pt,
               tightenlines,fleqn]{revtex4} %for preprint

\usepackage{amsmath,amssymb,revsymb,graphicx,dcolumn}
\usepackage{leftidx}
\usepackage{slashed}

\newcommand{\cev}[1]{\reflectbox{\ensuremath{\vec{\reflectbox{\ensuremath{#1}}}}}}
\newcommand{\beq}{\begin{equation}}
\newcommand{\eeq}{\end{equation}}
\newcommand{\beqa}{\begin{eqnarray}}
\newcommand{\eeqa}{\end{eqnarray}}
\newcommand{\bsubeqs}{\begin{subequations}}
\newcommand{\esubeqs}{\end{subequations}}

\begin{document}
%\noindent arXiv:1603.01148 
\hfill KA--TP--08--2016\vspace*{8mm}\newline
\title[]
      {Effective photon mass from black-hole formation \vspace*{5mm}}
\author{Slava Emelyanov}
\email{viacheslav.emelyanov@kit.edu}
\affiliation{Institute for Theoretical Physics,\\
Karlsruhe Institute of Technology (KIT),\\
76131 Karlsruhe, Germany\\}

\begin{abstract}
\vspace*{2.5mm}\noindent
We compute the value of effective photon mass $m_\gamma$ at one-loop level in QED in the background of
small ($10^{10}\,\text{g} \lesssim M \ll 10^{16}\,\text{g}$) spherically symmetric black hole in asymptotically
flat spacetime. This effect is associated with the modification of electron/positron propagator in presence of
event horizon. Physical manifestations of black-hole environment are compared with those of hot neutral
plasma. We estimate the distance to the nearest black hole from the upper bound on $m_\gamma$ obtained
in the Coulomb-law test. We also find that corrections to electron mass $m_e$ and fine structure constant
$\alpha$ at one-loop level in QED are negligible in the weak gravity regime.
\end{abstract}


\maketitle

\section{Introduction}

In Minkowski space the photon acquires an effective (thermal) mass if it propagates through a (neutral)
plasma of electrons and positrons held at high enough temperature, i.e. $T \gg m_e$. The effective
photon mass $m_\gamma$ turns out to be proportional to the temperature $T$ of the electron-positron
plasma at one-loop level~\cite{Weldon1} (see also~\cite{LeBellac,Kapusta&Gale}). This effect is
exponentially suppressed if the plasma is cold, i.e. $T \ll m_e$~\cite{Woloshyn,Nieves&Pal&Unger,Gies}.
This occurs because most of electrons and positrons are in the ground state at low temperature. This
leads in turn to suppression of  photon-electron and photon-positron scattering events with respect
to the photon-photon scattering which is the higher-loop effect~\cite{Tarrach}. In summary, we have
\beqa\label{eq:epmm}
m_\gamma^2 &\approx&  e^2T^2
\left\{
\begin{array}{cl}
1/6\,, & T \;\gg\; m_e\,, \\[2mm]
4(m_e\beta)^{\frac{1}{2}}\,\exp(-m_e\beta)/(2\pi)^{\frac{3}{2}}\,, & T \;\ll\; m_e\,
\end{array}
\right.
\eeqa
at one-loop approximation in quantum electrodynamics (QED), where $\beta \equiv 1/T$ is the inverse
temperature.

There is a comparably recent idea of the assignment of readings of the macroscopic thermometer with
the so-called Wick squared operator~\cite{Buchholz&Ojima&Roos,Buchholz}. This is also known as 
the local temperature operator. Specifically, if one treats a scalar field model with the conformal coupling
to gravity, then one can find that $\langle \hat{\Phi}^2(x)\rangle = T^2/12$ in a thermal state characterized
by the temperature $T$. This was also generalized and treated in curved spacetimes~\cite{Buchholz&Schlemmer,Solveen,Emelyanov1,Rabochaya&Zerbini}. 
Certain applications in flat space were studied in~\cite{Buchholz&Verch}.

Considering a scalar non-interacting field model with mass $m = m_e$, one can obtain
\beqa
\langle \hat{\Phi}^2(x)\rangle &\approx& \frac{1}{2}T^2
\left\{
\begin{array}{cl}
1/6\,, & T \;\gg\; m_e\,, \\[2mm]
2(m_e\beta)^{\frac{1}{2}}\,\exp(-m_e\beta)/(2\pi)^{\frac{3}{2}}\,, & T \;\ll\; m_e\,,
\end{array}
\right.
\eeqa
for the renormalized value of the Wick squared operator in the thermal state described by the temperature
$T$. Thus, a quantitative discrepancy arises between the effective photon mass squared and
$\langle \hat{\Phi}^2(x)\rangle$ at low temperatures. In fact, these quantities are diverse both physically
and mathematically. 
 
Nevertheless, it is tempting to conjecture that $m_\gamma^2 \propto \langle \hat{\Phi}^2(x)\rangle$ holds 
\emph{qualitatively} at high temperatures at the $\alpha$-order approximation in quantum electrodynamics. 
If one takes this relation for granted, then one can predict that the photon acquires an effective mass, for 
instance, in the background of small Schwarzschild black holes, i.e. $T_H \gg m_e$ and $M \gg M_\text{Pl}$, 
where $T_H = M_\text{Pl}^2/(8\pi M)$ is the Hawking temperature and $M_\text{Pl} = (\hbar c/G)^{\frac{1}{2}}$ 
is the Planck mass. For these black holes, the size of the event horizon is 
$r_H = 2MG/c^2 \ll 3{\times}10^{-14}\,\text{m}$.\footnote{We assume such black holes exist in nature which 
were formed through gravitational collapse under extreme conditions present in early universe~\cite{Hawking}.}

Indeed, if we consider eternal Schwarzschild geometry with a black hole of mass $M$, then physical vacuum
corresponds to the Hartle-Hawking state. Far away from the black hole, i.e. $r \gg r_H$, we have 
$\langle \hat{\Phi}^2(x)\rangle_H \approx T_H^2/12$ for the scalar field model conformally coupled to 
gravity~\cite{Candelas}. If the black hole has formed through the gravitational collapse, then one might expect 
the photon possesses an effective mass decreasing with distance as $1/r$, because 
$\langle \hat{\Phi}^2(x)\rangle_U \propto T_H^2(2M/r)^2$ in the Unruh state~\cite{Candelas}. Remarkably, we 
have recently surmised this dependence of $m_\gamma$ on the distance to the black hole from a different 
perspective~\cite{Emelyanov2}.

In this paper we analytically derive the effective photon mass $m_\gamma$ at one-loop level in QED in 
asymptotically flat spacetime with a small spherically symmetric black hole
($M_\text{Pl} \ll M \ll 10^{21}M_\text{Pl}$). We find that the above relation between the effective photon 
mass $m_\gamma$ and the expectation value 
of the Wick squared operator $\langle \hat{\Phi}^2(x)\rangle$ \emph{does} qualitatively hold in the 
high-temperature limit, i.e. $T_H \gg m_e$, for the Hartle-Hawking and Unruh state. 

We shall also show that the analogy between the hot plasma and the environment of a small black
hole formed through the gravitational collapse is incomplete in that the black-hole environment cannot support
plasmon-like and plasmino-like excitations. However, a point-like electric charge can be partially screened due
to the modification of the electric permittivity and the magnetic permeability of the vacuum in the black-hole
background.

Throughout this paper the fundamental constants are set to unity, $c = G = k_\text{B} = \hbar = 1$, unless
stated otherwise.

\section{Effective photon mass}
\label{sec:effective photon mass}

To compute the photon self-energy at one-loop level in the background of evaporating Schwarzschild black 
hole, we need to have the free fermion propagator $S(x,x')$. Since the Dirac equation 
$\big(i\slashed{\nabla} - m_e\big)\psi(x) = 0$ can also be written as $\big(\Box + m_e^2\big)\psi(x) = 0$,
it is enough, however, to deduce the scalar propagator $G(x,x')$. Indeed, the propagator $S(x,x')$ can 
then be obtained by acting on $G(x,x')$ by the operator $i\slashed{\nabla} + m_e$ (e.g., see~\cite{Birrell&Davies}).

\subsection{Scalar Wightman function}

As pointed out above, we need to compute the scalar two-point function in spacetime with the
Schwarzschild black hole of mass $M$. We start with a massive scalar field
\beqa\label{eq:sfe}
\big(\Box + m_e^2\big)\Phi(x) &=& 0\,,
\eeqa
and look for positive frequency modes in the following form
\beqa\label{eq:modes}
\Phi_{kjm}(x) &=& \frac{1}{(4\pi\omega)^{\frac{1}{2}}}\frac{e^{-i\omega t}}{r}\,R_{kl}(r)Y_{lm}(\theta,\phi)\,,
\eeqa
where $\omega = (k^2 + m_e^2)^{\frac{1}{2}}$ and $Y_{lm}(\theta,\phi)$ are the spherical harmonics.
Substituting \eqref{eq:modes} in the scalar field equation \eqref{eq:sfe}, we obtain
\beqa\label{eq:rsfe}
\frac{d^2}{dr_*^2}R_{kl}(r) + f(r)\left(\frac{\omega^2}{f(r)}
-\frac{l(l+1)}{r^2} - m_e^2 + \frac{f'(r)}{r}\right)R_{kl}(r) &=& 0\,, \quad f(r) = 1- \frac{r_H}{r}\,,
\eeqa
where $r_* = r + r_H\ln(r/r_H - 1)$ is the Regge-Wheeler radial coordinate and the prime stands for differentiation
with respect to $r$. There are two types of radial modes, namely the ingoing and outgoing one. We denote these
as $\cev{R}_{\omega l}(r)$ for the ingoing modes and $\vec{R}_{\omega l}(r)$ for the outgoing modes.
The Wightman two-point function, e.g. in the Boulware (B) state, is then
\beqa
\langle \hat{\Phi}(x)\hat{\Phi}(x') \rangle_B &=& \sum_{lm}\int \frac{d\omega}{4\pi\omega}\,
\frac{e^{-i\omega\Delta{t}}}{rr'}
Y_{lm}(\Omega)Y_{lm}^*(\Omega')\Big(
\vec{R}_{\omega l}(r)\vec{R}_{\omega l}^*(r') +
\cev{R}_{\omega l}(r)\cev{R}_{\omega l}^*(r')
\Big),
\eeqa
where $\Delta{t} = t- t'$ by definition. The sum over $m$ can be performed and yields 
\beqa
\sum\limits_{m = - l}^{m = +l}Y_{lm}(\Omega)Y_{lm}^*(\Omega') &=& \frac{2l+1}{4\pi}\,P_l(\cos\Theta)\,,
\eeqa
where $\cos\Theta \equiv \cos\theta\cos\theta' + \sin\theta\sin\theta'\cos(\phi - \phi')$ and $P_n(x)$ is the Legendre
polynomial.

It is hardly possible to solve the radial mode equation~\eqref{eq:rsfe} analytically, but one can always do that
numerically. However, employing results of~\cite{DeWitt,Candelas,Page}, we obtain in the limit of vanishing 
mass of the scalar field ($m_e \rightarrow 0$) that
\beqa\nonumber\label{eq:kvec}
\vec{K}_\omega({\bf{x}},{{\bf{x}}}') &\equiv& \frac{1}{4\pi rr'}
\sum\limits_{l = 0}^{+\infty}(2l + 1)\vec{R}_{\omega l}(r)\vec{R}_{\omega l}^*(r')P_l(\cos\Theta) 
\\[2mm]
&\approx& 
\frac{\Delta^{\frac{1}{2}}(\rho)\sin(\omega\rho)}{4\pi\omega\rho(f(r)f(r'))^{\frac{1}{2}}}
\left\{
\begin{array}{ll}
4\omega^2 - \frac{(f(r)f(r'))^{\frac{1}{2}}}{rr'}\,\Gamma_\omega\,, & r \rightarrow 2M\,, \\[3mm]
\frac{(f(r)f(r'))^{\frac{1}{2}}}{rr'}\,\Gamma_\omega\,, & r \gg 2M\,,
\end{array}
\right.
\eeqa
and
\beqa\nonumber\label{eq:kcev}
\cev{K}_\omega({\bf{x}},{{\bf{x}}}') &\equiv& \frac{1}{4\pi rr'}
\sum\limits_{l = 0}^{+\infty}(2l + 1)\cev{R}_{\omega l}(r)\cev{R}_{\omega l}^*(r')P_l(\cos\Theta) 
\\[2mm]
&\approx& 
\frac{\Delta^{\frac{1}{2}}(\rho)\sin(\omega\rho)}{4\pi\omega\rho(f(r)f(r'))^{\frac{1}{2}}}
\left\{
\begin{array}{ll}
\frac{(f(r)f(r'))^{\frac{1}{2}}}{rr'}\,\Gamma_\omega\,, & r \rightarrow 2M\,, \\[3mm]
4\omega^2 - \frac{(f(r)f(r'))^{\frac{1}{2}}}{rr'}\,\Gamma_\omega\,, & r \gg 2M\,,
\end{array}
\right.
\eeqa
where $\rho \equiv (2\sigma({\bf{x}},{{\bf{x}}}'))^{\frac{1}{2}}$, $\sigma({\bf{x}},{{\bf{x}}}')$ is the three-dimensional
geodetic interval for the ultrastatic or optical metric $\bar{g}_{\mu\nu} = g_{\mu\nu}/f(r)$, $\Delta({\bf{x}},{{\bf{x}}}')$
is the Van Vleck determinant~\cite{DeWitt65} and 
\beqa
\Gamma_\omega &\equiv& \sum_{l = 0}^{+\infty}(2l + 1)|B_{\omega l}|^2 \;\approx\; 27\omega^2M^2
\eeqa
in the DeWitt approximation~\cite{DeWitt}.

The scalar two-point function in the case when the outgoing and ingoing modes are ``heated up" to inverse
temperatures $\beta_1$ and $\beta_2$, respectively, is
\beqa
W_{\beta_1,\beta_2}(x,x') &=& \vec{W}_{\beta_1}(x,x') + \cev{W}_{\beta_2}(x,x') 
\\[5mm]\nonumber
&\approx&
\int\limits_0^{+\infty}d\omega
\left(\frac{\cos\left(\omega\Delta{t} + 
i\frac{\omega\beta_1}{2}\right)}{4\pi\omega\sinh\left(\frac{\beta_1\omega}{2}\right)}\vec{K}_\omega({\bf{x}},{{\bf{x}}}') + 
\frac{\cos\left(\omega\Delta{t} + i\frac{\omega\beta_2}{2}\right)}{4\pi\omega\sinh\left(\frac{\beta_2\omega}{2}\right)}
\cev{K}_\omega({\bf{x}},{{\bf{x}}}')\right).
\eeqa
The Hartle-Hawking state corresponds to $\beta_1 = \beta_2 = \beta = 2\pi/\kappa$, where $\kappa$ is a value of 
the surface gravity on the horizon $r = r_H$~\cite{Hartle&Hawking}. The Boulware state follows from the
Hartle-Hawking one if we set $\beta_1 = \beta_2 = +\infty$~\cite{Boulware}. The physical state for the black holes
formed through the gravitational collapse is known as the Unruh state~\cite{Unruh}. This corresponds to 
$\beta_1 = \beta$ and $\beta_2 = +\infty$.

We now define the commutator function
\beqa\label{eq:sc}
C(x,x') &=& \vec{C}(x,x') + \cev{C}(x,x')\,
\eeqa
which will be used below, where
\bsubeqs
\beqa
\vec{C}(x,x') &=& \vec{W}_{\beta_1}(x,x') - \vec{W}_{\beta_1}(x',x)
\;\approx\;
\int\limits_0^{+\infty}\frac{d\omega}{4\pi\omega}
\big(e^{-i\omega\Delta{t}} - e^{+i\omega\Delta{t}}\big)\vec{K}_\omega({\bf{x}},{{\bf{x}}}')\,,
\\[1mm]
\cev{C}(x,x') &=& \cev{W}_{\beta_2}(x,x') - \cev{W}_{\beta_2}(x',x)
\;\approx\;
\int\limits_0^{+\infty}\frac{d\omega}{4\pi\omega}
\big(e^{-i\omega\Delta{t}} - e^{+i\omega\Delta{t}}\big)\cev{K}_\omega({\bf{x}},{{\bf{x}}}')\,.
\eeqa
\esubeqs
It is worth noting that the commutator functions defined in the above manner do not depend on the temperatures.\footnote{This is true for non-interacting theories or in the leading order approximation of the perturbation theory.} 
In general, this occurs because the commutator of the field operators plays a role of the algebraic structure of the algebra of local field operators. This structure is independent on Fock space representations of the algebra. 
Hence, it remains the same, for instance, independent on whether one treats the Boulware or Unruh state.

\subsection{Spinor Feynman propagator}

In general, fermion anti-commutation function $C_f(x,x')$ is related with the scalar commutator function
as follows
\beqa
C_f(x,x') &=& \vec{C}_f(x,x') + \cev{C}_f(x,x') 
\;=\; \big(i\slashed{\nabla} + m_e\big)C(x,x')\,,
\eeqa
where
\bsubeqs
\beqa
\vec{C}_f(x,x') &=& \vec{S}_{\beta_1}(x,x') + \vec{S}_{\beta_1}(x',x)\,,
\\[1mm]
\cev{C}_f(x,x') &=& \cev{S}_{\beta_2}(x,x') + \cev{S}_{\beta_2}(x',x)\,.
\eeqa
\esubeqs

To compute one-loop contribution to the photon self-energy, one needs to find the Feynman propagator 
$S(x,x')$. This can be expressed through the anti-commutation function~\cite{Dolan&Jackiw}. 
Specifically, we have
\beqa\label{eq:ffp}
S(\omega|\bf{x},\bf{x}') &=& \int\frac{d\omega'}{2\pi}
\frac{iC_f(\omega'|\bf{x},\bf{x}')}{\omega-\omega' + i\varepsilon}
-n_{\beta_1}(\omega)\vec{C}_f(\omega|{\bf{x}},{\bf{x}}')
-n_{\beta_2}(\omega)\cev{C}_f(\omega|{\bf{x}},{\bf{x}}')\,,
\eeqa
where $\varepsilon \rightarrow +0$, the integral is over all $\omega'$ lying in $\mathbf{R}$,
\beqa
n_\beta(\omega) &=& \frac{1}{e^{\beta\omega} +1}\,,
\eeqa
and $C_f(\omega|\bf{x},\bf{x}')$ is the Fourier transform over time of the anti-commutator function, i.e.
\beqa\nonumber
C_f(\omega|\bf{x},\bf{x}') &\equiv& \int{d}\Delta{t}\,e^{+i\omega\Delta{t}}C_f(x,x')
\;=\; \int{d}\Delta{t}\,e^{+i\omega\Delta{t}}\big(i\slashed{\nabla}_x + m_e\big)C(x,x')\,,
\eeqa
where $C(x,x')$ is the scalar commutator given in~\eqref{eq:sc}.

\subsection{Photon self-energy at one-loop level}

\begin{figure}[t]
\includegraphics[width=6.5cm]{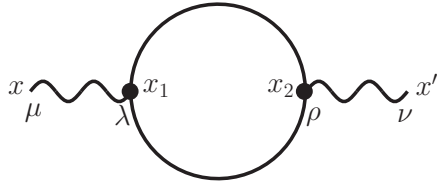}
\caption{One-loop vacuum polarization diagram.}
\end{figure}

In terms of the photon Feynman propagator, we have up to the $\alpha^2$-order in the perturbation
theory
\beqa\nonumber
G_{\alpha}^{\mu\nu}(x,x') &=& G^{\mu\nu}(x,x') - 4\pi\alpha\int{d}x_1dx_2(g(x_1)g(x_2))^{\frac{1}{2}}\,
\\[1mm]
&&\times\;G^{\mu\lambda}(x,x_1)\;
\text{Tr}\Big(\gamma_\lambda S(x_1,x_2)\gamma_\rho S(x_2,x_1)\Big)\;
G^{\rho\nu}(x_2,x') + \text{O}\big(\alpha^2\big)\,,
\eeqa
where we expect up to the $\alpha^2$-order that
\beqa
\Big(\Box_x\delta_{\lambda}^{\mu} + \big(m_\gamma^2\big)_{\lambda}^{\mu}\Big)G_{\alpha}^{\lambda\nu}(x,x')
&=& \frac{ig^{\mu\nu}(x)}{(-g(x))^{\frac{1}{2}}}\,\delta\big(x-x'\big)\,,
\eeqa
and $G^{\mu\nu}(x,x')$ satisfies this equation at the zeroth order in the fine structure constant $\alpha$. 
To find the effective photon mass squared, we thus need to compute
\beqa\label{eq:epm}
\big(m_\gamma^2\big)_{\lambda}^{\mu}G^{\lambda\nu}(\omega|{\bf{x}},{\bf{x}}') &=& -4\pi\alpha
\int d{\bf{y}}\sqrt{-g({\bf{y}})}\,K_\lambda^\mu(\omega,{\bf{x}},{\bf{y}})G^{\lambda\nu}(\omega|{\bf{y}},{\bf{x}}')
+ \text{O}\big(\alpha^2\big)\,,
\eeqa
where by definition
\beqa\label{eq:ik}
K_\lambda^\mu(\omega,{\bf{x}},{\bf{y}}) &= & \int\frac{d\Omega}{2\pi}\;
\text{Tr}\Big(\gamma^{\mu}S(\Omega|{\bf{x}},{\bf{y}})\gamma_{\lambda}S(\Omega-\omega|{\bf{y}},{\bf{x}})\Big)\,.
\eeqa

Since the black hole is small, we are working in the high-temperature limit. Consequently, one is allowed to use the 
hard thermal loop approximation~\cite{LeBellac,Kapusta&Gale}. In other words, we omit the electron mass $m_e$ 
in the fermion correlation function as well as the frequency $\omega$ as being negligible with respect to the temperature parameter $T_H$. Therefore, it is legitimate to employ the correlation function found above in the 
limit $m_e \rightarrow 0$ to obtain the effective photon mass in the temperature regime $T_H \gg m_e$ and 
$T_H \gg \omega$.

%Note that a neutral matter field is not asymptotically free when one takes gravity into consideration.
%This happens to be the case because gravitons are massless, which means the gravitational
%potential decreases too slowly for large $r$ to be neglected even at $r \rightarrow \infty$. This effect is
%similar to that in, e.g., quantum electrodynamics. To have free dynamics at spatial infinity one needs
%to define a physical matter operator which is covariant and possesses a correct gravitational potential.
%This can be achieved through dressing the Lagrangian scalar field operator by a cloud of the soft gravitons
%(as one does it in QED). The dressing should be universal for any field in the sense that local dynamics 
%is oblivious to that not to violate the equivalence principle. Therefore, the $1/r$-term in the radial 
%equation being due to $m_e^2 \neq 0$ does not influence local dynamics far away from the black hole.
%A detailed discussion of the dresssing of matter fields by the soft gravitons will be presented 
%in~\cite{Emelyanov&Queiruga}.\footnote{Note that the IR divergence in QED due to the masslessness
%of the photon does not show up in our treatment, because the vacuum polarisation diagram depicted
%in fig. 1 is insensitive to that.}

\subparagraph{Minkowski space}

One can employ the Feynman propagator given in~\eqref{eq:ffp} to compute the effective photon mass
at one-loop level in the high-temperature limit. This is achieved through setting $\beta_1 = \beta_2 = 1/T < \infty$
and substituting $M = 0$ in~\eqref{eq:kvec} and \eqref{eq:kcev}. The vanishing mass of the black hole 
implies that $\Gamma_\omega = 0$, because $\Gamma_\omega \approx 27\omega^2M^2$ in the DeWitt 
approximation~\cite{DeWitt}.

Our computation of the photon mass in the hot plasma will be non-standard if we work in the spherical 
coordinates. Nevertheless, we still have the result $m_\gamma^2 = e^2T^2/6$ as the theory is covariant.

In Minkowski space we can express~\eqref{eq:epm} and~\eqref{eq:ik} through the Cartesian coordinates and 
then perform the standard evaluations of the integrals. However, we can do the same when $M \neq 0$ far away
from the event horizon $r \gg r_H$. Comparing then the right-hand sides of~\eqref{eq:epm} in Minkowski space
and Schwarzschild space far from the black hole, we can immediately obtain the photon mass $m_\gamma$
due to the black hole in the weak gravity regime.

\subparagraph{Schwarzschild space}

The scalar propagator can be computed exactly in Minkowski space. This is not the case in Schwarzschild 
space even in the limit of the vanishing electron mass $m_e$. Our expressions for $\vec{K}_\omega({\bf{x}},{{\bf{x}}}')$ 
and $\cev{K}_\omega({\bf{x}},{{\bf{x}}}')$ given in~\eqref{eq:kvec} and \eqref{eq:kcev}, respectively, are reliable
whenever the points ${\bf x}$ and ${\bf x'}$ are close to each other.

The physical idea now is to notice that although one must integrate in~\eqref{eq:epm} over all values of ${\bf y}$, 
the main contribution to the integral will be from the spacial region in the vicinity of the point ${\bf x}$. In other 
words, the virtual electron-positron pair depicted in fig.~1 is a short-time or local event in spacetime.

This can also be exemplified by the computation of the photon self-energy in-between conducting plates in
the Casimir set-up. The photon propagator in-between the conducting plates differs from that in Minkowski 
space due to the non-trivial boundary conditions satisfied by the electromagnetic four-potential operator on 
the plates. The two-loop contribution to the photon self-energy after renormalization is then non-trivial, 
because of the internal photon propagator in the loops~\cite{Scharnhorst}. In the coordinate representation 
of the loop integrals, one would need to integrate over the whole spacetime. However, the main contribution 
will be when the vertices are close to each other. For instance, the result will be independent on the contributions 
from the points outside of the plates, where the photon propagator differs from that in-between the plates. 
This is fully consistent with computations based on the effective action of the electromagnetic field 
(fermion degrees of freedom are integrated out) at one-loop level in which one merely needs to have the 
propagator at space-time points being close to each other~\cite{Barton}.

This argument can also be supported by the empirical observations in the particle physics. Indeed, we have been
successfully employing the Minkowski-space approximation in studying various processes in the particle colliders.
However, the universe is non-flat at cosmological scales. According to the equivalence principle, nevertheless,
there always exists a local Minkowski frame. Therefore, the description of the scattering processes in QFT is
performed in the local Minkowski frame as if it is of the infinite extent. This is an adequate approximation
whenever relevant physics is characterised by a length scale being much smaller than a characteristic curvature
scale. In our case, this length is $l_c = R (R/r_H)^\frac{1}{2}$, where $R$ is the distance to the black-hole centre. 

Thus, we find that the (massless) scalar Feynman propagator in momentum space far away from the black-hole
horizon is approximatelly given by
\beqa\label{eq:up}
G_U(k,k') &\approx& \left(\frac{i}{k^2 + i\varepsilon} + 2\pi\left(\frac{27M^2}{4R^2}\right)n_\beta(\omega)
\delta\big(k^2\big)\right) \delta\big(k - k'\big)
\eeqa
at $T_H \gg m_e$ and $T_H \gg \omega$ in the Unruh state, where $k = (\omega,{\bf k})$. It is worth noticing that
$G_U(k,k')$ reduces to the ordinary scalar propagator in Minkowski space in the limit of the vanishing black-hole
mass $M \rightarrow 0$ or $R \rightarrow \infty$. This does not happen to be the case for the eternal black hole
described by the Hartle-Hawking state.

Having derived the propagator~\eqref{eq:up}, we can now obtain $m_\gamma$ at the one-loop approximation. 
For the black hole formed through the gravitational collapse, we find
\beqa\label{eq:epmf}
\left.m_\gamma^2\right|^\text{electron-positron} &\approx& \frac{27\pi\alpha}{24}\,T_H^2\left(\frac{r_H}{R}\right)^2 +
\text{O}\left(\frac{T_Hr_H^2}{R^3}\right)
\eeqa
far from the hole ($R \gg r_H$) in the high-temperature limit ($T_H \gg m_e$ and 
$M \gg M_\text{Pl}$).\footnote{The leading-order correction to the first term in~\eqref{eq:epmf} is due to the 
action of $\slashed{\nabla}$ on the prefactor $r_H^2/R^2$. The next-to-leading term is of the order of $r_H^3/R^4$.}
The formula~\eqref{eq:epmf} is the main result of our paper. It is worth noticing that $m_\gamma$ does not directly
depend on the black-hole mass $M$ at one-loop level in the leading order of the weak gravity limit.\footnote{Note 
that we have taken into account only electron-positron virtual pair to the photon self-energy. 
For instance, the same result holds for the virtual muon-antimuon pair, but then the black hole should be smaller
$T_H \gg m_\mu \gg m_e$ for not having exponentially suppressed contribution of this pair.
If $T_H \gg m_\mu$ holds, then $m_\gamma^2$ is $2$ times larger.} This does not mean $m_\gamma \neq 0$
for $M = 0$, because Eq.~\eqref{eq:epmf} has been computed under the assumption $M \neq 0$. Taking
the limit $M \rightarrow 0$ and evaluating the integral~\eqref{eq:ik} are not commuting operations. It is worth
reminding that the renormalised stress tensor depends on $M$ as $1/M^2R^2$, so that it is non-vanishing in
the limit $M \rightarrow 0$.

For the Hartle-Hawking state we obtain the standard result for $m_\gamma$ far away from the event horizon
like in the hot physical plasma in Minkowski space. Hence, small eternal black holes would considerably
influence photon kinematics. This is not a problem, because these black holes are not realizable through the
gravitational collapse anyway.

In the Standard Model the electromagnetic field corresponds to the $U(1)$ gauge group under which the vacuum
is invariant. This group is a subgroup of the spontaneously broken electroweak symmetry $SU(2)_L\times U(1)_Y$
with the electroweak phase transition occuring at the electroweak energy scale $M_\text{EW} \approx 10^2\;\text{GeV}$.
The temperature parameter $T_H$ for the small black holes is greater than $M_\text{EW}$ if the black-hole mass
$M \lesssim 10^{15}M_\text{Pl}$. It is not obvious whether one can rely on~\eqref{eq:epmf} when the black-hole mass
is smaller than $10^{15}M_\text{Pl}$. However, we expect that~\eqref{eq:epmf} is still reliable at least far from the
black hole, because all physical parameters are then small (see below). It is still not excluded that the phase
transition may occur at the distance $r_{ew} \approx 10^{-19}\,\text{m}$ for black holes of mass $M$ in the range
$M_\text{Pl} \ll M \ll 10^{16}M_\text{Pl}$. Note that the phase transition should occur far from the horizon as
$r_H \ll r_{ew}$. A similar observation was made long ago in~\cite{Hawking81,Moss85}.

We have focused on the black holes of mass $M \ll 10^{16}\,\text{g}$ for which $T_H \gg m_e$ and implicitly
presumed that the quasi-equilibrium approximation holds, i.e. spacetime is quasi-static. For sufficiently small black
holes, this approximation does not hold, because of the black-hole evaporation. This implies that $m_\gamma$
given in Eq.\eqref{eq:epmf} should be a reliable result for $M$ in the range
\beqa
10^{10}\,\text{g} &\lesssim& M \;\ll\; 10^{16}\,\text{g}\,,
\eeqa
where we have chosen the lower bound on $M$ by requiring that the smallest black hole has at least a
one-day lifetime (assuming the evaporation lasts till the complete disappearance of the black hole).
This range corresponds to $3{\times}10^{14}\,\text{g} \lesssim M_0 \ll 10^{16}\,\text{g}$ of the initial
mass of the primordial black holes.

\section{Concluding remarks}
\label{sec:concluding remarks}

The space-time structure significantly modifies when a black hole forms. The algebraic structure of a set of
the quantum field operators also modifies. As a consequence, propagators of quantum fields have a different
form in comparison with that in Minkowski space. Far away from the black holes, one might expect that 
quantum field theory becomes almost indistinguishable from its formulation in Minkowski spacetime. Indeed,
quantum field theory formulated in Minkowski space is well tested and verified in the particle colliders, 
although there are a lot of gravitational sources in our universe which make the geometry of spacetime be
of non-Minkowskian form at sufficiently large length scales.

Although it is legitimate to await of recovering Minkowskian quantum field theory far away from the black holes,
there must be specific imprints of these gravitational sources in physical experiments performed on earth.
In this paper, we have investigated these imprints of the small spherically symmetric black hole 
($10^{10}\,\text{g} \lesssim M \ll 10^{16}\,\text{g}$) on the effective photon mass at the $\alpha$-order 
approximation in QED.
Physically, it might be a consequence of the event-horizon formation which leads to the modification of the
quantum field operators (as the field equations explicitly depend on the black-hole mass $M$). Assuming
the process of the black-hole formation is unitary, the total quantum system (gravity and matter fields) evolves
semi-classically if the backreaction of the quantum fields on the geometry is small. This is usually described
by saying that the quantum fields occupy the Unruh state~\cite{Unruh}. It inevitably implies the presence of
the thermal-like correction in Eq.\eqref{eq:up} yielding $m_\gamma \neq 0$.\footnote{In a hot physical
electron-positron plasma in Minkowski space, the thermal photon mass is a result of the photon interaction
with the plasma particles. We could interpret $m_\gamma \neq 0$ as due to a hot plasma of Hawking
electrons and Hawking positrons, but then we should accept a phase transition during black-hole formation.
In fact, these particles are elements of the non-Minkwoskian Fock space representation of the field operator
algebra far away from the black hole, where one expects to have the ordinary representation~\cite{Emelyanov}.}

The Wick squared operator cannot always be interpreted as a macroscopic temperature 
squared~\cite{Emelyanov1}. Moreover, the effective photon mass $m_\gamma$ for the Boulware state vanishes. 
However, the Wick squared $\langle \hat{\Phi}^2(x)\rangle_B$ is non-zero and negative. Specifically, 
$\langle \hat{\Phi}^2(x)\rangle_B \propto - T_H^2(2M/r)^4$ as this can be shown employing the Page 
approximation~\cite{Page}. Therefore, the qualitative validity of the relation between $m_\gamma^2$ and
$\langle \hat{\Phi}^2(x)\rangle$ is counter-intuitive for this state. It should be noted, however, that all divergencies in
the evaluation of $m_\gamma$ has been subtracted, such that $m_\gamma$ vanishes in the Boulware state.
The result of this renormalization is finite and depends on the parameter $T_H$. This is completely analogous
to that of how one proceeds in the hot plasma in flat space. The Wick squared operator in turn also depends on
how one renormalizes it. One usually defines this operator as follows
\beqa\nonumber
\hat{\Phi}^2(x) &=& \lim_{x' \rightarrow x}\big(\hat{\Phi}(x)\hat{\Phi}(x') - H(x,x')\hat{1}\big)\,,
\eeqa
where $H(x,x')$ is the Hadamard parametrix. This definition is state-independent. It is worth noting that the
Wick squared operator can also be written down as $\hat{\Phi}^2(x) = {:}\hat{\Phi}(x)\hat{\Phi}(x){:}$, where
colons refer to the normal order product.

We have recently found in~\cite{Emelyanov2} that the two-loop or, possibly, even higher-loop effect is dominant
far from the small black holes if the wave-length $\lambda_\gamma$ of the electromagnetic radiation is in the
range $\lambda_e \ll \lambda_\gamma \ll \alpha^{\frac{1}{2}}\lambda_e(T_H/m_e)$, where $\lambda_e$ is the
Compton length of the electron. However, the one-loop dominance occurs whenever 
$\alpha^{\frac{1}{2}}\lambda_e(T_H/m_e) \ll \lambda_\gamma \ll l_c$, where $l_c$ is a characteristic 
curvature scale.

\subsection{Plasma-like environment of black hole}

Quantum fluctuations of the electromagnetic and fermion field around the black holes reveal plasma-like 
properties. This can be characterised by the modification of the electric permittivity $\epsilon(\omega,k,R)$ and
the magnetic permeability $\mu(\omega,k,R)$ of the vacuum. This is analogous to a similar phenomenon in the 
Casimir set-up~\cite{Scharnhorst,Barton}. Note that $\epsilon(\omega,k,R) = 1/\mu(\omega,k,R)$ in the limit
$M \rightarrow 0$ as in the Minkowski vacuum, because the second term in~\eqref{eq:up} vanishes.
The same holds for $M \neq 0$, but in the spatial infinity, i.e. at $R \rightarrow \infty$.

The normal hot plasma is characterised by two parameters, namely $\alpha$ and temperature $T_H$.
The black-hole plasma-like environment is described by one more parameter which is of the order of 
$T_H(2M/R)$.  Although the temperature parameter $T_H$ is large with respect to $m_e$, the 
plasma-like environment is ``cold" in the sense that the plasma-like frequency $\omega_p$ is small for 
$R \gg r_H$ with respect to $T_H$, i.e.
\beqa
\omega_p &\approx& \left(\frac{27\pi\alpha}{36}\right)^{\frac{1}{2}}T_H \left(\frac{r_H}{R}\right)
\;\ll\; T_H\,.
\eeqa

The photon propagator has the longitudinal and transverse part in the hot plasma~\cite{Weldon1}.
The longitudinal part becomes a propagating degree of freedom (a collective mode mediated by the plasma 
particles) known as plasmon for frequencies $\omega_l \sim \omega_p$, while the transverse part, photon,
is dynamical for $\omega_t \geq \omega_p$. In the black-hole background, it implies that there should exist
a plasmon-like excitation. However, the plasmon wavelength $\lambda_{p} \sim 10^2\,R$ is much bigger than
$R$. In general, our approximation is only reliable for $\lambda_\gamma \ll R$ as pointed out above. Thus, 
there are no plasmon-like waves in the black-hole background.\footnote{It might be an effect of
the absence of the plasmon's mediators. This appears to be in agreement with~\cite{Emelyanov}.}

The physical plasma is opaque for electromagnetic waves with frequency $\omega$ less than the plasma 
frequency. Thus, these electromagnetic waves are reflected due to the collective response of the plasma particles.
The poles in the photon propagator also disappear at $\omega < \omega_p$ close to the black hole. 
For instance, we find that the region $r_H \ll R \lesssim 1\,\text{nm}$ should be opaque for the light 
wave of length $\lambda_\gamma = 500\,\text{nm}$. Therefore, we have $\lambda_\gamma \gg R$. In the 
hot plasma of the size $R$, one would expect merely a negligible damping of the wave amplitude. 
We expect the same effect for the small black holes due to the non-trivial manifestation of vacuum 
fluctuations. The reflection of the light waves from the plasma-like environment of the black hole should 
be an extremely rare event (if at all).

The plasma frequency in the normal plasma is a classical quantity, i.e. that does not explicitly depend on the
Planck constant $\hbar$. Indeed, one has $\omega_p^2 = 4\pi e^2 n/m_e$ in the cold plasma, 
where $n$ is a density number of the particles~\cite{Jackson,Ginzburg}. In the hot physical plasma
$n \sim T^3$ and $m_e \sim T$ resulting in $\omega_p^2  \sim e^2T^2$. In particular, we have 
$\omega_p^2 = e^2T^2/9$ for the neutral electron-positron plasma (e.g., see~\cite{LeBellac}). However, 
our result~\eqref{eq:epmf} cannot be understood classically, because of the quantum 
nature of the parameter $T_H$ ($\propto \hbar$).

Moreover, the plasma-like environment of small black holes far away from the horizon is effectively characterised 
by a new (local) temperature parameter
\beqa
T_L &=& \frac{3\sqrt{3}}{16\pi}\,M_\text{Pl}\,\frac{L_\text{Pl}}{R}\,
\eeqa
which is much smaller than $T_H$ far from the hole ($R \gg r_H$), where $L_\text{Pl} = (\hbar G/c^3)^{\frac{1}{2}}$
is the Planck length. The numerical factor can deviate from its exact value as we have been working in
the DeWitt approximation. The same scaling of the temperature from the
distance has been recently found in~\cite{Brustein&Medved} within a different framework.

\subsection{Modified Coulomb law}

One can employ our formula~\eqref{eq:epmf} to estimate the distance to the nearest small black hole 
from the upper bound on the photon mass. In the hot physical plasma, Coulomb's potential of a point-like electric 
charge is exponentially suppressed far from the charge as $\exp(-r/r_D)$, where 
$r_D = 1/m_D = (\sqrt{2}m_\gamma)^{-1}$ 
is the Debye radius. This phenomenon is known as the Debye screening (e.g., see~\cite{LeBellac,Kapusta&Gale}
and~\cite{Jackson,Ginzburg} in the case of the hot and cold plasma, respectively).
In our situation, the electromagnetic field effectively becomes a short range interaction. 

We obtain from $m_\gamma \lesssim 10^{-14}\,\text{eV}$~\cite{Williams&Faller&Hill} that the small
black hole at that time could not be closer to the laboratory than $R$, where
\beqa
R &\approx& 8.6{\times}10^5\;R_\odot \left(\frac{10^{-18}\,\text{eV}}{m_\gamma}\right) \;\gtrsim\; 250\;\text{km}\,,
\eeqa
where $R_\odot \approx 2.95\,\text{km}$ is sun's gravitational radius. Neglecting any other possible contributions to the effective photon mass, then the stronger upper bound on $m_\gamma$, the farther small black 
hole should be from the laboratory.

It is worth emphasizing that the Debye screening of the charge due to the black hole cannot be complete
(within our approximation), because the Debye radius is much bigger than $R$, specifically $r_D \gtrsim 8.8{\times}10^4\,\text{km}$. The fact $R \ll r_D$ does not imply our approximation is unreliable. Indeed, the conducting shell used 
in~\cite{Williams&Faller&Hill} to test the Coulomb law has a size about $1\,\text{m}$ which is much smaller 
than the distance to the black hole $R$.

In the physical plasma the Debye screening occurs due to the collective response of the plasma particles to 
the external electric charge. In our case, it is a vacuum polarization effect. In the absence of the black 
hole or very far away from it, the photon is almost massless at any order of the perturbation theory due to 
the gauge and Lorentz symmetry. Not too far from the black hole, spacetime isometry starts to significantly
deviate from the Minkowskian one due to the black-hole horizon. As a consequence, the vacuum response to
the electromagnetic field operator described by the electric permittivity and magnetic permeability modifies. 
This eventually results in the non-trivial photon dispersion relation. A similar effect occurs in-between the 
conducting plates, wherein, however, (low-energy) photons remain massless~\cite{Scharnhorst,Barton}.

\subsection{One-loop correction to electron mass $m_e$ and fine structure constant $\alpha$}

The electron mass is also modified in the black-hole background. Following~\cite{Weldon2,Levinson&Boal}
(see also~\cite{LeBellac,Kapusta&Gale}), we obtain at one-loop level that
\beqa
\delta m_e &\approx& \left(\frac{27\pi\alpha}{32}\right)^{\frac{1}{2}}T_H \left(\frac{r_H}{R}\right)
\eeqa
The correction to $m_e$ is thus negligibly small with respect to $m_e$ if $R \gg 4.4{\times}10^{-15}\,\text{m}$. 
It is worth mentioning 
that classical estimate of the electron size is about $2.8{\times}10^{-15}\,\text{m}$. In the hot plasma, the thermal
correction $\delta{m}_e$ to the electron mass is much bigger than $m_e$. In our case, this correction to $m_e$ 
is suppressed by the factor $r_H/R$. Hence, we have $m_e \gg \delta{m}_e$, although $T_H \gg m_e$. 
This implies there are no plasmino-like excitations in the background of the small evaporating black holes. This is fully 
consistent with of having no mediator due to which these collective modes could propagate. 

The temperature-dependent correction to the fine structure constant $\alpha$ in the hot plasma has been
derived in~\cite{Donoghue&Holstein&Robinett}. In the background of the small black hole we find
\beqa
\alpha(M) &\approx& \alpha\left(1 
+ \frac{2\alpha}{3\pi}\left(\frac{27r_H^2}{16R^2}\right)\ln\left(\frac{M_\text{Pl}^2}{8\pi Mm_e}\right)\right).
\eeqa
The effective fine structure constant $\alpha(M)$ approaches $\alpha$ in the limit $M \rightarrow 0$.
Its maximal value in the range $M_\text{Pl} \ll M \ll 10^{21}M_\text{Pl}$ slightly differs from $\alpha$. 
Specifically, the deviation of $\alpha(M)$ from $\alpha$ is much smaller than $10^{-8}$ for those values of 
the black-hole mass. At the distance $1\,\text{m}$ from the black hole, this deviation becomes $10^{-15}$ 
times smaller.

\subsection{Black holes in analogue gravity}

The effect we have derived in this paper is due to the interaction between photons and electrons/positrons
and the presence of the small black hole.
In the $\lambda\Phi^4$-model, the massless scalar particle acquires an effective mass $m_\Phi$ in the
background of the black holes as well. Following~\cite{LeBellac}, one can obtain
\beqa
m_\Phi^2 &\approx& \frac{27\lambda}{384}\,T_H^2\left(\frac{r_H}{R}\right)^2
\eeqa
at one-loop level far away from the event horizon. We shall treat this theory in a forthcoming
paper~\cite{Emelyanov3} near evaporating black holes, where one may expect a breakdown of the
perturbation theory analogous to that observed in~\cite{Emelyanov2}.

There is an analogue of black holes in a medium known as a dumb hole~\cite{Unruh2} (see also~\cite{Liberati}
for a comprehensive review of analogue gravity). Experimental evidences have been recently reported
in favour of the dumb-hole evaporation which is supposed to be analogous to the black-hole
evaporation~\cite{Weinfurtner,Steinhauer}.

For fluids in which phonons are self-interacting, one might expect a non-trivial dispersion relation for the 
phonon similar to that for the photon far from the small black hole. Specifically, an effective phonon mass
might depend on the distance to the sonic horizon.

\section*{%\hspace*{-4.5mm}
ACKNOWLEDGMENTS}

It is a pleasure to thank Frans Klinkhamer for discussions, references and many important comments 
on an early version of this paper from which I benefited a lot. We are grateful to Frasher Loshaj and 
Jos\'{e} Queiruga for discussions and comments. We would like also to thank Ramy Brustein for
an interesting discussion.

\end{document}